\documentclass[aps,twocolumn,showpacs,superscriptaddress]{revtex4}

\usepackage{graphicx}
\usepackage{latexsym}

\begin{document}

\title{Dynamical Diagram and Scaling in Polymer Driven Translocation}

\author{Takuya Saito}
\affiliation{Department of Physics, Kyushu University 33, Fukuoka 812-8581, Japan}

\author{Takahiro Sakaue\footnote{Corresponding author: sakaue@phys.kyushu-u.ac.jp}}
\affiliation{Department of Physics, Kyushu University 33, Fukuoka 812-8581, Japan \\ PRESTO, Japan Science and Technology Agency (JST), 4-1-8 Honcho Kawaguchi, Saitama 332-0012, Japan}

\def\degC{\kern-.2em\r{}\kern-.3em C}

\def\SimIneA{\hspace{0.3em}\raisebox{0.4ex}{$<$}\hspace{-0.75em}\raisebox{-.7ex}{$\sim$}\hspace{0.3em}} 

\def\SimIneB{\hspace{0.3em}\raisebox{0.4ex}{$>$}\hspace{-0.75em}\raisebox{-.7ex}{$\sim$}\hspace{0.3em}}

\date{\today}

\begin{abstract}
By analyzing the real space nonequilibrium dynamics of polymers, we elucidate the physics of driven translocation and propose its dynamical scaling scenario analogous to that in the surface growth phenomena. 
We provide a detailed account of the previously proposed tension-propagation formulation and extend it to cover the broader parameter space relevant to real experiments; In addition to a near-equilibrium regime, we identify three distinct nonequilibrium regimes reflecting the steady-state property of a dragged polymer with the finite extensibility.
Finite size effects are also pointed out.
These elements are shown to be crucial for the appropriate comparison with experiments and simulations.
\end{abstract}


\maketitle

\section{Introduction}

Macromolecules can pass through a nanoscopic hole which is much smaller in space dimension than the bulk coil size.
Given the chain-like architecture, this fact itself may not be surprising, but the way how it goes threading is interesting.
Today, this mode of molecular transport, called translocation, finds its great potentiality in the multidisciplinary field of nano-technology and biological sciences~\cite{PNAS_Kasianowicz_1996,PRL_Henrickson_2000,NanoLett_Storm_2005,PRE_Storm_2005,BJ_Wanunu_2008,JPhys_Milchev_2011}, e.g., in inventing nano-pore based rapid DNA sequencing devices.

In typical situations, a polymer (with $N_0$ segments) passes through driven by the force $f$ acting at the pore site.
The corresponding setup has been realized in most of experiments, where the charged polymers (DNAs) are driven by the electric field, and the voltage drop is concentrated in the pore~\cite{PNAS_Kasianowicz_1996,PRL_Henrickson_2000,NanoLett_Storm_2005,PRE_Storm_2005,BJ_Wanunu_2008}.
From the technological design perspective, one of the most important measure is the translocation time $\tau(N_0, f)$, i.e., an average passage time for a polymer to go through.
For this reason, there have been considerable experimental attempts to quantify the translocation time.~\cite{PNAS_Kasianowicz_1996,PRL_Henrickson_2000,NanoLett_Storm_2005,PRE_Storm_2005,BJ_Wanunu_2008}.
Extensive numerical simulations have also revealed a rich variety of the dynamics~\cite{PRE_Luo_2008,EPL_Luo_2009,EPL_Lehtola_2009,EPL_Dubbeldam_2007,JPhys_Vocks_Panja_2008,PRE_Bhattacharya_Binder_2010,PRE_Fyta_2008}.
Under certain conditions, i.e., when the specific pore-polymer interactions can be neglected, these results could be described in the light of the universality inherent in the chain-like molecules~\cite{PRL_Sung_1996,PRE_Chuang_2001,PRE_Kantor_2004,PRE_Sakaue_2007,PRE_Sakaue_2010}. 
It is not, however, clear how and to what extent this is possible. 
Diverse ``apparent" scaling laws of the translocation time reported in literatures remain to be explained~\cite{NanoLett_Storm_2005,PRE_Storm_2005,BJ_Wanunu_2008,JPhys_Milchev_2011,PRE_Luo_2008,EPL_Luo_2009,EPL_Lehtola_2009,EPL_Dubbeldam_2007,JPhys_Vocks_Panja_2008,PRE_Bhattacharya_Binder_2010,PRE_Fyta_2008,PRL_Sung_1996,PRE_Chuang_2001,PRE_Kantor_2004}.

\begin{figure}
\begin{center}
\includegraphics{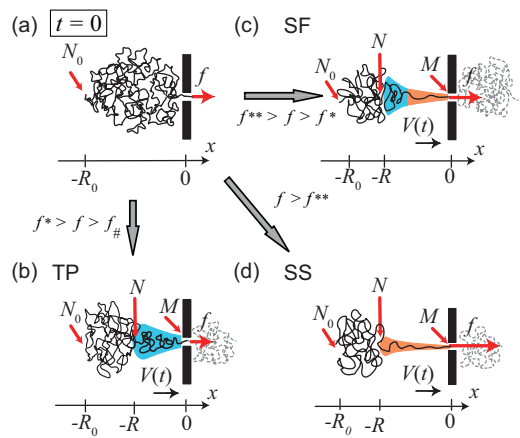}
      \caption{
	  (Color Online) Illustration of various conformations during the driven translocation process starting from (a) initial equilibrium conformation. (b) Trumpet [TP]. (c) Stem-flower [SF]. (d) Strong-stretching [SS]. More complete processes in respective regimes are shown in Figs.~5-7. The tensed ``moving domain" is shaded, while the translocated part is represented by a dotted curve. 
	  }
\label{fig1}
\end{center}
\end{figure}

In this paper, we find three distinct nonequilibrium regimes with the different exponents through the tension-propagation mechanism proposed in \cite{PRE_Sakaue_2007,PRE_Sakaue_2010}.
The characteristic nonequilibrium conformations (trumpet [TP], stem-flower [SF], strong-stretching [SS]) of the polymeric chain with the finite extensibility are shown in Fig.~1, through which we establish a dynamical scaling scenario covering various models/situations and broad parameter regions relevant to real experiments.

\begin{figure}
\begin{center}
\includegraphics{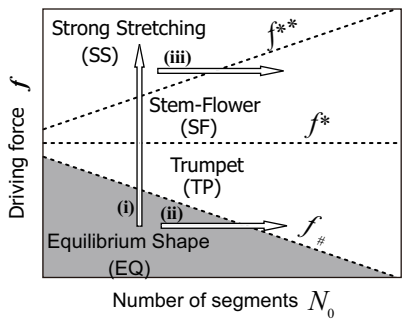}
      \caption{ Dynamical diagram of the driven translocation process in the plane of $N_0$ and $f$ (double logarithmic scale). Shaded region represents the case to which the near-equilibrium theory applies~\cite{NanoLett_Storm_2005,PRE_Kantor_2004,PRE_Sakaue_2010}.
The arrows (i)-(iii) indicate typical sweeping paths of parameters in experiments (see Figs.~\ref{fig3} and~\ref{fig4}).
The boundaries $f_\sharp \simeq N_0^{-\nu}$, $f^* \simeq 1$, $f^{**} \simeq N_0^{\nu}$ separate regimes characterized by different conformations (Fig.~1).
}
\label{fig2}
\end{center}
\end{figure}

\section{Transient Response}
We begin by clarifying the exponents relevant to equilibrium (static) and near-equilibrium (dynamic) situations. First, the Flory exponent $\nu$ characterizes the fractality of the equilibrium polymer coil whose size is given by $R_0 \simeq a N_0^{\nu}$ ($a$: segment size).
Equally important is the dynamical exponent $z$ characterizing the dynamics of ``critical fluctuation" (polymer coil)~\cite{deGennesBook}. 
Normally $z=3$ in dilute solutions (non-draining), but the case with $z=(1+2\nu)/\nu$ (free draining or Rouse dynamics) is also conceptually important. The latter indeed corresponds to models in most (but not all) of numerical simulations. 
By denoting the translocation time $\tau \sim N_0^\alpha f^{-\beta}$, our aim is to find these dynamical exponents $\alpha$, $\beta$ relevant to the driven translocation in terms of $\nu$ and $z$.

There is a growing evidence that a key element lies in the dynamical property of the polymer; it cannot respond to the driving force all at once. This becomes evident for rather weak forces $ f \ge f_{\sharp} \simeq (k_{\rm B}T/a)N_0^{-\nu}$, i.e., the thermal energy $k_{\rm B}T$ divided by the equilibrium coil size $R_0$.
In what follows, we make the length, force and time dimensionless by the units $a$, $k_{\rm B}T/a$ and $\eta a^3/(k_{\rm B}T)$, respectively, with $\eta$ being the viscosity of the solvent.
To analyze the case $f> f_{\sharp}$, let us see the sketches of a translocating polymer where the driving force acts on the segment at the pore ($x=0$) as shown in Fig.~1. Note that $x$-axis is taken to be perpendicular to the wall with the pore. The one chain end is first sucked at time $t=0$, while the polymer keeps the equilibrium conformation in $x \leq 0$.
By the time $t$, $M(t)$ segments have been already translocated, and the subsequent $N(t)-M(t)$ segments in the anterior {\it cis} side (shaded in Fig.~1) 
are set in motion under the tension, while the remaining $N_0 -N(t)$ segments in the rear part are still at rest.

\section{Two phase picture}

Although plausible, a mathematical formulation of the above picture is nontrivial. In fact, at any moment, each segment moves with its own velocity and the complete characterization of the transient process would require handling highly nonlinear equations in multi-dimensional space with excluded-volume/hydrodynamic interactions. A drastic simplification is achieved, however, by introducing the `` two phase picture", i.e., a moving and a quiescent domains separated by the tension front, and applying the steady-state approximation for the former (Fig.~1). It amounts to say that the relevant properties of the moving domain can be captured by the representative velocity $V(t)$ and the steady-state conformation of the $N(t)-M(t)$ segments dragged at its velocity. Leaving its justification in the Appendix I, we remark here that there remains only single gross variable, i.e., the position of the front, or the boundary between the moving and the quiescent domains.
This coarse-graining makes the problem analytically treatable.
 
Then, by considering the mass conservation relation, we can show that the boundary position obeys the following equation;
\begin{eqnarray}
\sigma (- R) \left( \frac{ {\rm d}R(t)}{ {\rm d}t } + V(t)  \right)
=
\frac{ {\rm d} N(t)}{ {\rm d} t },
\label{mass_free}
\end{eqnarray}
where $\sigma(x)$ is the line density of the segment and $x=-R(t)$ is the location of the boundary (see Appendix II for this derivation). Note that we first consider only the tension-propagation stage (Fig.~1\,(b),\,(c),\,(d)) since this is the dominant process giving the scaling exponents in the translocation.

In addition, $N(t)$-th segment is just located in the boundary during the moving domain growth and then it still retains an initial rest position.
Thus, by averaging over the initial conformation at $t=0$, we find the relation
\begin{equation}
N(t)^{\nu}
\simeq
R(t).
\label{front_1}
\end{equation}

\section{Steady-state of a Dragged Polymer}

Here we write the {\it dynamical equation of state} relating the steady-state velocity $V$, the extension $R$ and the driving force $f$ in the form;
\begin{eqnarray}
VR= {\mathcal J}(f).
\label{D_ef_state}
\end{eqnarray}
The scaling form ${\mathcal J} (f) \simeq f^{p}$ and the segment line density at the boundary 
\begin{eqnarray}
\sigma(-R) \simeq V^{-q}
\label{sigma_R_Vq}
\end{eqnarray}
 are expressed through the exponents:
\begin{eqnarray}
p =\left\{
           \begin{array}{ll}
              z-2 &   \qquad(f_{\sharp}< f<f^*)   \\
             1 &  \qquad (f >f^*)  
           \end{array}
        \right.
        \label{D_ef_state_J}
\end{eqnarray}
\begin{eqnarray}
q =\left\{
           \begin{array}{ll}
              \frac{(1-\nu)}{(z-1)\nu} &   \qquad( f_{\sharp}<f<f^{**}) \\
              0 &  \qquad (f >f^{**}) 
           \end{array}
        \right.
        \label{sigma_R}
\end{eqnarray}
As already mentioned, $f_\sharp \simeq N_0^{-\nu}$ signifies the onset of the nonequilibrium dynamics. Moreover, we find two characteristic forces $f^* \simeq 1$, $f^{**} \simeq N_0^\nu$. This leads to the identification of three distinct nonequilibrium regimes on the dynamical diagram as summarized in Fig.~2. 
The derivation and the physical meaning of eqs.~(\ref{D_ef_state})-(\ref{sigma_R}) and boundaries will be clarified in the following subsections concerning each regime.

{\it Trumpet {\rm (TP)} regime}:

Let us first consider the situation with the driving force in the range $f_{\sharp}< f < f^* \simeq 1$. Here the blob concepts enable one to handle the space dependent length scale, whereby provide a generic description of the deformed dragged polymer irrespective of the model details~\cite{PRE_Sakaue_2010,EPL_Brochard_1993} (Fig.~\ref{fig1}\,(b)). Evaluating the dragging force and the elastic restoring force in terms of the blob size $\xi(x) \simeq g(x)^{\nu}$ at each point, we have the local force balance equation as follows:
\begin{eqnarray}
 V \int_{-R}^{x} {\rm d}x' \, \xi(x')^{z-3} \simeq [ \xi(x)]^{-1}  \Leftrightarrow 
 \xi(x)^{2-z}
\simeq
V (R + x).~~
\label{semi_local_force}
\end{eqnarray}
Substituting $x=0$ combined with the total force balance $f \simeq 1/\xi( 0)$, we obtain the dynamical equation of state eq.~(\ref{D_ef_state}) with eq.~(\ref{D_ef_state_J}) (top), which corresponds to a generalization of the trumpet model for a tethered polymer under flow~\cite{EPL_Brochard_1993}.
At the boundary, there is the largest blob of size $\xi_R$ centered around $x = -R + \xi_R$. Then, eq.~(\ref{semi_local_force}) yields 
\begin{equation}
\xi_R \simeq V^{1/(1-z)},
\label{rear_blob}
\end{equation}
which leads to the segment line density $\sigma(- R) \simeq g_R/\xi_R \Rightarrow$ eq.~(\ref{sigma_R}) (top) with eq.~(\ref{sigma_R_Vq}).

{\it Strong stretching {\rm (SS)} regime}:

At the upper threshold of the Trumpet regime $f =f^* \simeq 1$, the number of segment in the smallest blob at the pore vicinity becomes very small $g(0)\simeq 1 \Leftrightarrow \xi(0) \equiv \xi^* \simeq 1$. For stronger forces, the blob description is no longer valid, and one has to take account of the finite chain extensibility of the backbone (e.g., Fig.~\ref{fig1}\,(c),\,(d)).
The line density such highly stretched parts is estimated as $\sigma(x) \simeq 1$ at the scaling level, which indicates eq.~(\ref{sigma_R}) (bottom) with eq.~(\ref{sigma_R_Vq}).
Let us consider the limiting case of the strong driving so that the moving domain of the chain is highly stretched close to the full extension $R(t) \simeq a(N(t)-M(t))$~\cite{EPL_Brochard_1995} in the entire stage of the translocation (Fig.~\ref{fig1}\,(d)). This is realized when $f > f^{**} \simeq N_0^{\nu}$~\footnote{From eq.~(\ref{D_ef_state}) and eq.~(\ref{D_ef_state_J})\,(bottom), the velocity is estimated as $V \simeq f/R_0 \simeq f N_0^{-\nu}$ when the tension arrives at another chain end in the SS regime, so that the drag force acting on the rear end segment is $V\simeq f N_0^{-\nu}$ ($\Leftrightarrow \eta a V \simeq \eta a^2 N_0^{-\nu}/k_{\rm B}T$ in the original unit) at that moment. The SS shape moreover requires $V > f^*$ to retain the strong stretching even at rear end segment. Eventually, by combining it with $V \simeq f N_0^{-\nu}$, we find the lower bound as $f > N_0^{\nu} \simeq f^{**}$ for SS regime. }. 
Neglecting the logarithmic correction due to the hydrodynamic interactions, 
the total drag force is given by $f \simeq (N(t)-M(t))V$, and then the scaling form of the dynamical equation of motion is specified by eq.~(\ref{D_ef_state_J}) (bottom) irrespective of the solvent property (free-draining or non-draining).

{\it Stem-flower {\rm (SF)} regime}:

In the intermediate between above two regimes $f^{*} < f < f^{**}$, the moving domain takes a highly stretched conformation only in the early stage, for which the analysis in the SS regime applies.
This point will be discussed in the section~VI again.
The moving domain grows with time, and the velocity $V$ decreases towards $V^* \simeq 1$, at which the tension at the boundary becomes comparable to $f^*$, that is a ``Blooming point", before which the moving domain composes of stem only. From then on, the conformation of the moving domain can be described as a stem-flower (Fig.~\ref{fig1}~(c)).
Compared to the SS regime, the chain in SF regime has smaller extension because of the coiling at the rear part.
Still, scaling exponents associated with the entire domain in the SF regime is indistinguishable from that in the SS regime
~\cite{EPL_Brochard_1995} and satisfies $R(t) \simeq a[N(t)-M(t)]$ and $f \simeq V [N(t)-M(t)]$. Consequently, we find eq.~(\ref{D_ef_state_J}) (bottom); its distinction with that in the SS regime lies in a smaller numerical prefactor, while the conformation at the rear can be described by the TP model (eq.~(\ref{sigma_R}) (top)).
Therefore, the SF regime shares (i) the dynamical equation of state with the SS regime, (ii) the segment line density at the boundary with the TP regime.

\begin{figure}
\begin{center}
\includegraphics{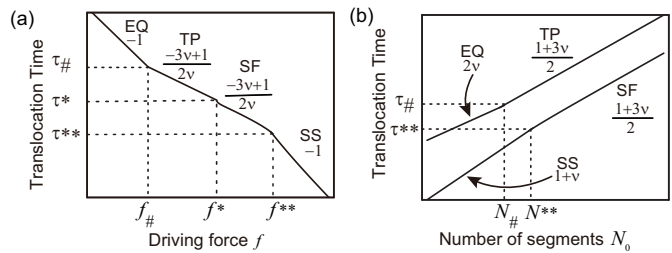}
      \caption{Plots of $\tau (N_0, f)$ in the non-draining case (double logarithmic scale). (a) and the upper/lower plots in (b) correspond to the parameter sweepings indicated by arrows (i) and (ii)/(iii) in Fig.~\ref{fig2}.}
\label{fig3}
\end{center}
\end{figure}
\begin{table}
\begin{tabular}{|c|c|c|}
\hline
 & \multicolumn{2}{ |c|}{ Non-draining }  \\ \cline{2-3}
  & $\nu=0.50$  &  $\nu = 0.588$  \\  \hline
EQ 	&   $\alpha=1.00$, $\beta=1.00$          & $\alpha=1.18$, $\beta=1.00$    \\ \hline
TP  &     $\alpha=1.25$, $\beta=0.50$   & $\alpha=1.38$, $\beta=0.65$  \\ \hline
SF	        &   $\alpha=1.25$, $\beta=0.50$      & $\alpha=1.38$, $\beta=0.65$  \\  \hline
SS    	&      $\alpha=1.50$, $\beta=1.00$     & $\alpha=1.59$, $\beta=1.00$    \\ 
\hline
\end{tabular}
\label{table1}
\caption{Predicted exponents $\alpha$ and $\beta$ in the non-draining case ($z=3$). Note that $\gamma = \nu / \alpha $.}
\end{table}

\section{Dynamical Scaling: Tension Propagation}
A set of basic eqs.~(\ref{mass_free})-(\ref{sigma_R_Vq}), combined with the appropriate steady state conformational properties (eqs.~(\ref{D_ef_state_J}) and~(\ref{sigma_R})), leads to a differential equation:
\begin{eqnarray}
[R^{(1-\nu)/\nu} - f^{-pq}R^{q}]\frac{{\rm d}R}{{\rm d}t} \simeq f^{p(1-q)}R^{q-1}.
\label{R_diff_eq}
\end{eqnarray}
This should be solved with the initial condition $R_{\rm ini}$ at $t= \tau_{\rm ini}$, which plays a role in the correction to scaling (see below).
Collecting the leading order terms, one obtains the dynamical scaling for the front evolution, which is asymptotically ($t \gg \tau_{\rm ini}$) valid, in the form
\begin{eqnarray}
R(t_f) \simeq t_f^{\gamma},
\end{eqnarray}
where the growth exponent $\gamma = \nu/(1+\nu-\nu q)$ characterizes the time evolution of the tension front and the force exponent $\beta = p(1-q)$ encodes the rescaling of time $t_f = t f^{\beta}$ with respect to the driving force. This stage persists until the tension reaches the rear end, from which the propagation time follows as 
\begin{equation}
\tau_{\rm p} \simeq N_0^{\alpha} f^{-\beta},
\label{tau_p}
\end{equation}
with $\alpha = \nu/\gamma$.
Notably, such a scaling relation among the roughness, growth and dynamical exponents characterizing the nonequilibrium tension propagation process is reminiscent to that in the surface growth phenomena~\cite{BarabasiBook}.

\begin{figure}
\begin{center}
\includegraphics{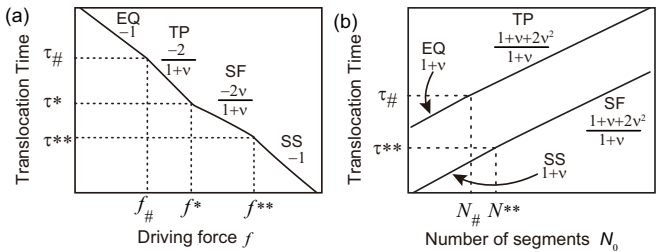}
      \caption{ The same plots as those in Fig.~\ref{fig3} in the free-draining case.}
\label{fig4}
\end{center}
\end{figure}
\begin{table}
\begin{tabular}{|c|c|c|}
\hline
 & \multicolumn{2}{ |c|}{ Free-draining }  \\ \cline{2-3}
   & $\nu =0.50$  & $\nu = 0.588$ \\  \hline
EQ	& $\alpha=1.50$, $\beta=1.00$    & $\alpha=1.59$, $\beta=1.00$     \\ \hline
TP 	        & $\alpha=1.33$, $\beta=1.33$   & $\alpha=1.43$, $\beta=1.26$ \\ \hline
SF	        & $\alpha=1.33$, $\beta=0.66$   & $\alpha=1.43$, $\beta=0.74$ \\ 
 \hline
SS	    & $\alpha=1.50$, $\beta= 1.00$     & $\alpha=1.59$, $\beta=1.00$     \\ 
\hline
\end{tabular}
\label{table2}
\caption{ The same as TABLE I in the free-draining case.}
\end{table}

Our main results are summarized in Figs.~\ref{fig3} and~\ref{fig4}, where we plot $\tau_{\rm p}(N_0, f)$ in the non-draining ($z=3$)  and the free-draining ($z=(1+2\nu)/\nu$) cases, respectively.
These are expected to be valid for long chains, and indeed capture a number of observations made in numerical/real experiments. We also listed the concrete values of exponents for cases $\nu=1/2$ and $\nu=\nu_3 \simeq 0.5876$ in tables~1 and~2.
As a primal message, we stress that the identification of the appropriate regime (depending on $N_0$ and $f$) as well as exponents $\nu$, $z$ characterizing the equilibrium polymer coil is of a crucial importance for the proper comparison with experiments.
For example, we first note that most simulations treat free-draining chains and the typical driving force is on the order of $f \simeq 1$ (TP and/or SF regimes), and the reported values of $\alpha$ exponent are scattered around $\sim 1.4$ in three dimension ($\nu=\nu_3$). In particular, recent high-accuracy simulation results $\alpha = 1.41 \pm 0.01$ (Langevin dynamics) and $\alpha = 1.42 \pm 0.01$ (atomistic molecular dynamic)~\cite{PRE_Luo_2008} are in excellent agreement with our prediction $\alpha = 1.43$. The change of $\alpha$ exponents with the increase in $f$ and/or with the inclusion of hydrodynamic interactions has been examined in ref.~\cite{EPL_Lehtola_2009}, whose qualitative trend accords with ours.
In experiments using double-stranded DNA~\cite{NanoLett_Storm_2005,BJ_Wanunu_2008}, the non-draining approximation with $\nu=1/2$ would be appropriate~\cite{deGennesBook}, and the typical driving force range $10~{\rm pN}$ would correspond to SF or even SS regime for short DNA chains. Then, the reported values $\alpha=1.27 \pm0.03$~\cite{NanoLett_Storm_2005} and $\alpha \simeq 1.4$~\cite{BJ_Wanunu_2008} are in good agreement with our prediction.
On the other hand, much less attention has been paid to the $\beta$ exponents. Usually, it has been implicitly assumed to be $\beta=1$, but a recent study has claimed $\beta \simeq 0.8$ for strong force~\cite{EPL_Luo_2009} in reasonable agreement with our prediction $\beta=0.74$ in SF regime ($\nu=\nu_3$, free-draining).

\begin{figure}
\begin{center}
\includegraphics{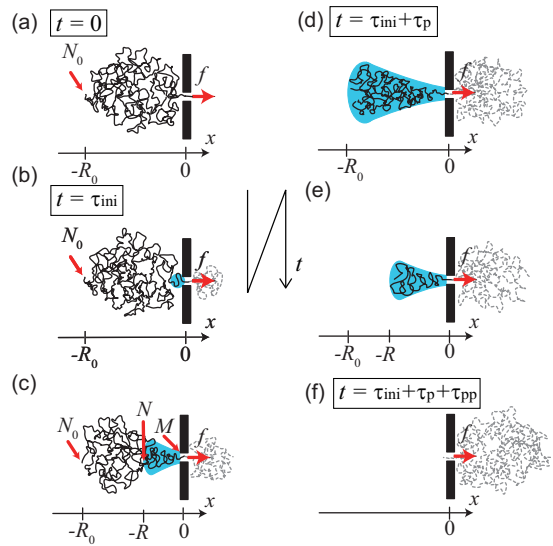}
      \caption{ (Color Online) The sketch of the entire process in trumpet (TP) regime. (a) Initial equilibrium conformation. (b) Formation of the initial tensed blob.
	  (c) Propagation stage of the tensile force retaining the TP shape. (d)/(e) Beginning/mid point of the post-propagation stage. (f) End of the translocation process. 
	  The tensed ``moving domain" is shaded, while the translocated part is represented by a dotted curve.
	  }
\label{fig5}
\end{center}
\end{figure}

\begin{figure}
\begin{center}
\includegraphics{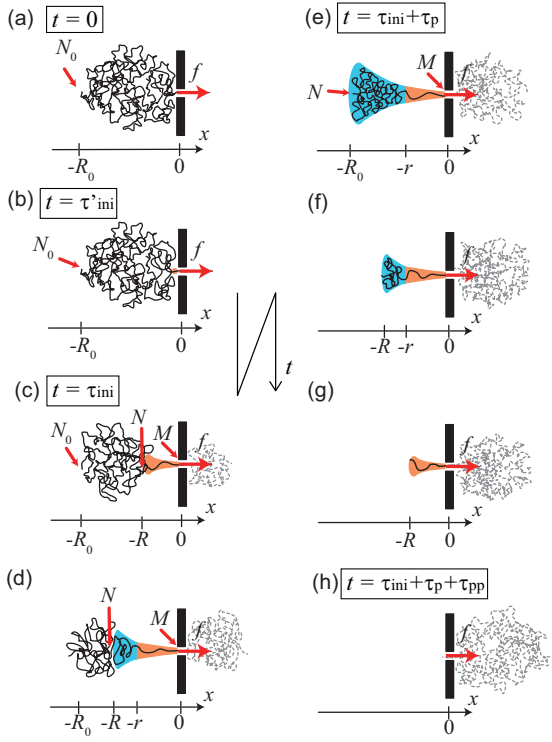}
      \caption{ (Color Online) The sketch of the entire process in stem-flower (SF) regime. (a) Initial equilibrium conformation. (b) Relaxation of first tensed segment. 
	  (c)/(d) Propagation stage of the tensile force; (c) ``blooming point" (end of the only stem conformation)/(d) stem-flower. (e) End of the propagation stage (beginning of the post-propagation stage). (f)/(g) the post-propagation with (e) stem-flower/(g) only stem. (h) End of the translocation process. The domains shaded by blue or orange correspond to flower or stem, respectively ($x=-r$ is the junction point between stem and flower). 
	  }
\label{fig6}
\end{center}
\end{figure}

\begin{figure}
\begin{center}
\includegraphics{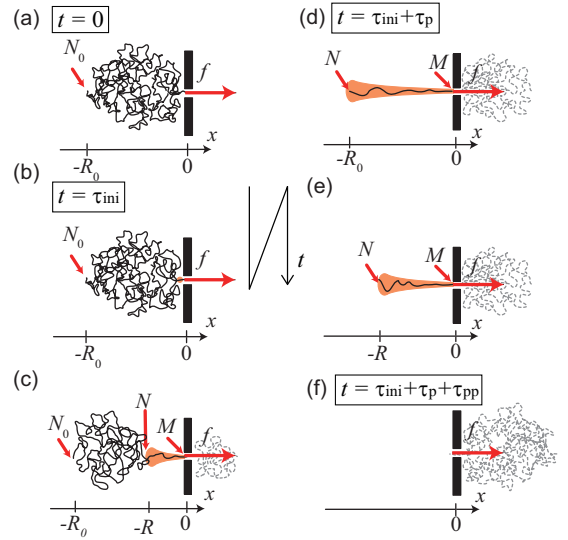}
      \caption{ (Color Online) The sketch of the entire process in strong-stretching (SS) regime. (a) Initial equilibrium conformation. (b) Relaxation of first tensed segment.
	  (c) Propagation stage of the tensile force. (d)/(e) Beginning/mid point of the post-propagation stage. (f) End of the translocation process. 
	  }
\label{fig7}
\end{center}
\end{figure}

\section{Finite Size Effects}
How can we go one step further beyond the leading term scaling argument?
Practically, resolving complications associated with various finite-size and/or crossover effects is an important issue, for which we point out two main sources.
The translocation time can be formally written as $\tau = \tau_{\rm ini}+ \tau_{\rm p} + \tau_{\rm pp}$.
Figures~5, 6 and 7 show the full story of the driven translocation process in TP, SF and SS regimes, respectively.
For long chains, the tension propagation stage (Fig.~5(b)\,$\rightarrow$(d), Fig.~6\,(c)$\rightarrow$(e), Fig.~7\,(b)$\rightarrow$(d)) dominates the whole process in the nonequilibrium driven regimes ($f>f_\sharp$), thus, $\tau_{\rm p}$ represents the translocation time $\tau \simeq \tau_{\rm p}$. For shorter chains, however, this might be severely affected by (i) the post-propagation stage ($\tau_{\rm pp}$) and/or (ii) the effect associated with the initial conditions in eq.~(\ref{R_diff_eq}) ($\tau_{\rm ini}$ and $R_{\rm ini}$).
For the point (i), we can describe the process after $t=\tau_{\rm ini}+\tau_{\rm p}$ by setting ${\rm d}N/{\rm d}t=0$ in eq.~(\ref{mass_free}) as follows:
\begin{equation}
-R \frac{{\rm d}R}{{\rm d}t} \simeq f^p, ~~~~~~~~(\tau_{\rm ini} +\tau_{\rm p} \leq t \leq \tau)
\end{equation}
where $R(t)$ is the location of the rear end and approaching to the hole (Fig.~5\,(d)$\rightarrow$(f), Fig.~6\,(e)$\rightarrow$(h), Fig.~7\,(d)$\rightarrow$(f)) with $R(\tau_{\rm ini}+\tau_{\rm p}) = R_0$.
The estimate for its characteristic duration time $\tau_{\rm pp}= \tau -(\tau_{\rm ini} + \tau_{\rm p})$ is obtained by setting $R(\tau)=0$, thus, 
\begin{equation}
\tau_{\rm pp} \simeq R_0^2 f^{-p} \simeq N_0^{2\nu} f^{-p}
\label{tau_pp}
\end{equation}
with the exponent $p$ given by eq.~(\ref{D_ef_state_J}).

The point (ii) can be taken into account by the following estimates;
\begin{eqnarray}
 R_{\rm ini} \simeq\left\{
           \begin{array}{ll}
              \xi_0 \simeq f^{-1} &   \qquad( f_{\sharp}<f<f^{*}) \\
              r(\tau_{\rm ini}) \simeq f &  \qquad (f^{*} < f <f^{**}) \\
              1 &  \qquad (f^{**} < f) 
           \end{array}
        \right.
\end{eqnarray}
\begin{eqnarray}
 \tau_{\rm ini} \simeq\left\{
           \begin{array}{ll}
              \xi_0^{z} \simeq f^{- z} &   \qquad(  f_{\sharp}<f<f^{*}) \\
              G^{\alpha_{\rm (SS)}}f^{-\beta_{\rm (SS)}} \simeq f^{1/\nu}&  \qquad (f^{*} < f <f^{**}) \\
              1 &  \qquad (f^{**} < f) 
           \end{array}
        \right.
		\label{tau_ini}
\end{eqnarray} 
The first one ($f_{\sharp}<f<f^*$) comes from the formation of the initial tensed blob of size $\xi_0 \simeq f^{-1}$ (Fig.~5\,(b)), which becomes the microscopic time $\sim 1$ at the threshold $f = f^*$. This microscopic time corresponds to $\tau_{\rm ini} \simeq 1$ in the SS regime as in Fig.~7\,(b).  
In the SF regime ($f^{*} < f <f^{**}$), we choose $\tau_{\rm ini}$ to be the blooming time, i.e., the moment when the flower part starts to appear (Fig.~6\,(c)). Before that,
there is only a stem part in the early period (Fig.~6\,(b) $\rightarrow$ (c)) during which the analysis for the SS regime applies after the microscopic relaxation time of first tensed segment $\tau_{\rm ini}' \sim 1$ as in Fig.~6\,(b). These lead to the time $\tau_{\rm ini} \simeq G^{\alpha_{\rm (SS)}}f^{-\beta_{\rm (SS)}}$ after which the coiled flower-conformation appears at the rear part ($\alpha_{(\rm ss)}$, $\beta_{(\rm ss)}$ are the exponents for SS regime). Here, by denoting the stem length as $r (t)$ (Fig.~6\,(d)), the pertinent segment number is $G \simeq [r (\tau_{\rm ini})]^{1/\nu}$ from eq.~(\ref{front_1}), where the stem length at the beginning of the late SF regime is $r(\tau_{\rm ini}) \simeq f$ (See Appendix III).
Solving eq.~(\ref{R_diff_eq}) with these initial conditions results in the shortening of $\tau_{\rm p}$. In particular, near the thresholds $f \rightarrow f^{**} (-0)$ and $f \rightarrow f_{\sharp} (+0)$, we find $R_{\rm ini} \rightarrow R_0$, thus, $\tau_{\rm p}$ become vanishingly small. 
In the former, $\tau_{\rm ini}$ is already close to $\tau_{\rm p} (f \rightarrow f^{**} (+0))$.
In the latter case, the contributions from $\tau_{\rm ini}$ and $\tau_{\rm pp}$ are comparable, and $\tau_{\rm pp}$ crossovers to $\tau_{\rm pp}^{\rm (EQ)} \simeq N_0^{(z-1)\nu}f^{-1}$, which is a result from the equilibrium shape assumption~\cite{PRE_Sakaue_2010}.
These factors may result in the narrowing of the TP regime and, in general, cause the rounding-off the crossover at $f_{\sharp}$ as well as $f^{**}$.

To summarize this section, one finds by comparing eqs.~(\ref{tau_p}), (\ref{tau_pp}) and (\ref{tau_ini}) that the propagation stage dominates, i.e., $\tau \simeq \tau_{\rm p}$ for large $N$ limit in driven translocation regime ($f>f_\sharp$). For real systems with small $N$, however, the finite-size correction might be apparent. In particular, $\tau_{\rm pp}$ may become dominant for short chains in two dimension, thus, an appreciable crossover scaling with respect to $N$ would be expected as pointed out in ref.~\cite{PRE_Sakaue_2010} . On the other hand, for weaker forces ($f < f_\sharp$), the equilibrium shape assumption would be valid, and $\tau \simeq \tau_{\rm pp}^{\rm (EQ)}$. We note, however, that in this weaker forces ($f <f_\sharp$), the segment returning through the pore becomes apparent, and this may alter the exponent in $\tau_{\rm pp}^{\rm (EQ)}$~\cite{JPhys_Vocks_Panja_2008}, while such an effect is negligible under the strong driving force.

\section{Remarks and Perspective}
Once again, we emphasize that the key lies in the nonequilibrium transient response property inherent in a long flexible chain. 
Our strategy to implement this is a two phase picture and applying a steady state approximation for the moving domain.
Before closing, let us make several remarks on the derived exponents.
(i) Exponents in the translocation time $\alpha$, $\beta$ reflect the exponent $p$, $q$ characterizing the steady state conformation of the dragged chain (eqs.~(\ref{D_ef_state})-(\ref{sigma_R})).
We have adopted eqs.~(\ref{D_ef_state_J}), (\ref{sigma_R}), which is correct in the scaling limit, for the purpose of the general discussion, but a care should be taken in the comparison with real experiments. 
In particular, the logarithmic correction due to the hydrodynamic interactions (in non-draining case) is neglected, and this might alter the effective exponent $p$, $q$ , thus, $\alpha$, $\beta$, too, in the certain range of the interest~\cite{Science_Perkins_1995,PRE_Larson_1997}. Although the results in ref.~\cite{PRE_Storm_2005,PRE_Fyta_2008} are in good agreement with our prediction, further careful checks are awaited on it.
(ii) As stated in Appendix I, the steady state approximation is by construction marginally valid to capture the salient features of the driven translocation, but there is some room to improve the theory by incorporating a slight deviation from the approximation. 
(iii) The presence of several different regimes and the finite-size effects makes the accurate determination of the exponents very difficult (see also the remark (i), (ii) above). We expect, however, that it would be possible to capture a global trend in the exponents by carefully analyzing numerical/experimental data obtained in widely different conditions.  
(iv) Aside from exponents, the proposed physical picture is generic, and we believe that our formalism constitutes a minimal model, or a prototype in the problem.  
From our point of view, it is more important to comprehend the underlying physics (even qualitatively), and seek for possible improvements of theory to describe the phenomena more adequately (see, for example, a recent attempt by Rowghanian and Grosberg~\cite{JPCB_Rowghanian_Grosberg_2011}). 
(v) One of the interesting predictions concerns the beta exponent; it is not necessarily unity in contrast to claims in most other works. We expect that such a trend (even qualitatively) could be detected by careful experiments; its reliable trend should serve as a stringent test for any theory of the driven translocation.

\section{Summary}

In summary, we proposed dynamical scaling scenario applicable to the broad range of situations met in the study of polymer driven translocation. 
The paper was devoted to the detailed account of tension-propagation mechanism proposed in~\cite{PRE_Sakaue_2007,PRE_Sakaue_2010} and its extension to cover the wider parameter space through the distinct nonequilibrium conformations. 
Also shown are the finite size effect, various associated crossovers and their rounding off. 
We believe that these elements operate as a guide for further studies towards comprehensive understanding of the driven translocation.
Last of all, we expect that the emergent physical picture may be relevant to many other polymeric systems driven by local and strong stimuli. Best examples would be found in the micro-manipulation experiments as well as the biopolymer functions in cells.

{\bf Appendix I.--- Validity of the steady-state approximation:} Let us look at the moving domain at some time during the tension propagating stage (e.g., the shaded part in Fig.~1\,(b)).

Since $N(t)$-th segment at the front have just taken part in the moving domain and the velocity profile would be smooth even around the front, there might be substantial velocity gradient there. 
If so, its magnitude can be estimated as $\dot{\gamma}_{\xi_R} \simeq V/\xi_R$, whose $\xi_R$ is the size of the largest blob at the front (eq.~(\ref{rear_blob})) and can be regarded as the boundary layer. Comparing the shear rate with the corresponding longest relaxation time $\tau_{\xi_R} \simeq \xi_R^z$, we find $\tau_{\xi_R} \dot{\gamma}_{\xi_R} \simeq V \xi_R^{z-1} \simeq 1 $ using eqs.~(\ref{rear_blob}). Thus the concerned shear rate is marginal and one can apply the steady state approximation.

As another possibility, one may suppose that an uniform velocity gradient is entirely spread in the moving domain with the shear rate given by $\dot{\gamma}_R \simeq V/R$. If so, the longest relaxation time in the entire tensed chain is $\tau_R \simeq R/V$, and we similarly have the relation as $\tau_R \dot{\gamma}_R \simeq 1 $~\cite{Macromol_Brochard_1995}. Here $\tau_R$ was estimated as follows; The dynamical equation of state (\ref{D_ef_state}) $f=VRf/{\mathcal J}(f)$ implies the spring constant $k_R= Vf/{\mathcal J}(f)$ and the viscous frictional coefficient $\Gamma_R = Rf/{\mathcal J}(f)$, which lead to the relaxation time $\tau_R \simeq \Gamma_R/k_R = R/V$. Although obtaining real velocity profile in the moving domain requires a more thorough analysis, which is beyond the scope of the present study, one may expect it to be something between the above two extremes. We can therefore expect the steady state approximation would be valid to address the salient features in the driven translocation.

{\bf Appendix II.---Mass conservation:} 
Here we discuss the derivation of eq.~(\ref{mass_free}) from the integral form of the mass conservation in the moving domain;
\begin{eqnarray}
N(t)-M(t) = \int_{-R(t)}^{0} \sigma(t,x) {\rm d}x.~~~~~~~~~~~({\rm A-1}) \nonumber
\label{total_mass}
\end{eqnarray}
Taking account of the moving boundary $R(t)$ associated with the tension propagation, the time derivative of eq.~(A-1) leads to 
\begin{eqnarray}
\frac{{\rm d}[N(t)-M(t)]}{{\rm d}t} &=& \int_{-R(t)}^{0} \frac{\partial \sigma(t,x)}{ \partial t} {\rm d}x + \sigma (t,-R) \frac{{\rm d}R}{{\rm d}t} \nonumber \\
							  &=& j (t,-R) -j(t,0) + \sigma (t,-R) \frac{{\rm d}R}{{\rm d}t}, \nonumber \\ 
							  &&~~~~~~~~~~~~~~~~~~~~~~~~~~~~~~~~~({\rm A-2})\nonumber
\label{mass_derivation}
\end{eqnarray}
where the continuity equation $\partial \sigma(t,x)/\partial t + \partial j(t,x)/\partial x=0$ is used in the last equality. 
The pore current in eq.~(A-2) would be strongly dependent on the pore specificity, but can be eliminated by its definition $j(t,0) = {\rm d}M/{\rm d}t$. 
Then, eq.~(\ref{mass_free}) follows by assuming $j(t,-R) \bigl(\equiv \sigma (t,-R) V (t,-R) \bigr) =\sigma (t,-R) V (t)$, where $V(t)$ is the steady state velocity of the moving domain.
Under this assumption, the number of the translocated segments $M(t)$ should be calculated from eq.~(A-1), not from time integration of $\sigma (0) V(t)$. 
The latter quantity does not satisfy the conservation of mass as $V(t)$ characterizing the chain deformation slightly differs from the velocity of the $M(t)$-th segment at the pore.
We note that a better estimate of the velocities $V(t,-R)$ at the boundary (due to the slight derivation from the representative velocity $V(t)$) would improve the result.

{\bf Appendix III.---Blooming point condition in the SF regime:}
At the end of the early SF regime (composed of stem only), the following conditions are met for the stem length $R$ and the moving velocity 
\begin{eqnarray}
\eta R V \simeq f, ~~~\eta a V \simeq \frac{k_{\rm B}T}{a}~~~~({\rm in~the~original~unit}).~~({\rm A-3}) \nonumber
\end{eqnarray}
These are respectively the total force balance and the force balance at the rear end with the blob of size $\sim a$.
From these, one obtains $V \simeq \frac{k_{\rm B}T}{\eta a^2}$, $R \simeq \frac{fa^2}{k_{\rm B}T}$ or in the dimensionless form $V \equiv V^* \simeq 1$, $R \equiv r(\tau_{\rm ini}) \simeq f$.


\begin{thebibliography}{}
	
\bibitem{PNAS_Kasianowicz_1996}
  {Kasianowicz J.~J., Brandin E., Branton D. and Deamer D.~W.,
  Proc. Natl. Acad. Sci. U.S.A. \textbf{93}, (1996) 13770.}
	
\bibitem{PRL_Henrickson_2000}
 Henrickson S.~E., Misakian M., Robertson B., and Kasianowicz J.~J.,
  Phys. Rev. Lett. \textbf{85}, (2000) 3057.
    
\bibitem{NanoLett_Storm_2005}
	Storm A.~J., Storm C., Chen J., Zandbergen H., Joanny J.-F., and Dekker C.,
    Nano Lett. \textbf{5}, (2005) 1193.  

\bibitem{PRE_Storm_2005}
	Storm A.~J., Chen J.~H., Zandbergen H.~W., and Dekker C.,
    Phys. Rev. E \textbf{71}, (2005) 051903.  
	
\bibitem{BJ_Wanunu_2008}
	Wanunu M., Sutin J., McNally B., Chow A., and Meller A.,
    Biophys. J. \textbf{95}, (2008) 4716.  
	
	
\bibitem{JPhys_Milchev_2011}
	Milchev A.,
    J. Phys.: Condens. Matter \textbf{23}, (2011) 103101.  
		
	
\bibitem{PRE_Luo_2008}
 	Luo K., Ollila S.~T.~T., Huopaniemi I., Ala-Nissila T., Pomorski P., Karttunen M., Ying S.-C., and Bhattacharya A.,
    Phys. Rev. E \textbf{78}, (2008) 050901.  
  
  
\bibitem{EPL_Luo_2009}
  	Luo K., Ala-Nissila T., Ying S.-C., and Metzler R.,
    Europys. Lett. \textbf{88}, (2009) 68006. 
	
  
\bibitem{EPL_Lehtola_2009}
    Lehtola V.~V., Linna R.~P., and Kanski K.,
    Europys. Lett. \textbf{85}, (2009) 58006. 

	
\bibitem{EPL_Dubbeldam_2007}
   Dubbeldam J.~L.~A., Milchev A., Rostiashvili V.~G. and Vilgis T.~A.,
    Europys. Lett. \textbf{79}, (2007) 18002.  
  
 
\bibitem{JPhys_Vocks_Panja_2008}
    Vocks H., Panja D., Barkema G.~T. and Ball R.~C.,
    J. Phys.: Condens. Matter \textbf{20}, (2008) 095224.   
  
\bibitem{PRE_Bhattacharya_Binder_2010}
     Bhattacharya A. and Binder K.,
    Phys. Rev. E \textbf{81}, (2010) 041804.

\bibitem{PRE_Fyta_2008}
  Fyta M., Melchinonna S., Succi S. and Efthimios K.,
   Phys. Rev. E \textbf{78}, (2008) 036704.   
	
\bibitem{PRL_Sung_1996}
    Sung W. and Park P.~J.,
    Phys. Rev. Lett. \textbf{77}, (1996) 783. 
  

\bibitem{PRE_Chuang_2001}
    Chuang J., Kantor Y. and Kardar M.
    Phys. Rev. E \textbf{65}, (2001) 011802. 	
	
	
\bibitem{PRE_Kantor_2004}
	Kantor Y. and Kardar M.,
    Phys. Rev. E \textbf{69}, (2004) 021806.
	
	
\bibitem{PRE_Sakaue_2007}
 	Sakaue T.,
    Phys. Rev. E \textbf{81}, (2007) 021803.


\bibitem{PRE_Sakaue_2010}	
 	Sakaue T.,
    Phys. Rev. E \textbf{81}, (2010) 041808. 
 
 
\bibitem{deGennesBook}
  de~Gennes P.~G.,
  \textit{Scaling Concepts in Polymer Physics}
  (Cornell University Press, Ithaca 1979).


  
\bibitem{EPL_Brochard_1993}
   	Brochard-Wyart F.,
    Europhys. Lett. \textbf{23}, (1993) 105. 

\bibitem{EPL_Brochard_1995}	
	 Brochard-Wyart F.,
    Europhys. Lett. \textbf{30}, (1995) 387. 

\bibitem{BarabasiBook}
  Barab${\rm \acute{a}}$si A.-L. and Stanley H. E.,
  \textit{Fractal Concepts in Surface Growth}
  (Cambridge University Press, New York 1995).

\bibitem{Science_Perkins_1995}
  Perkins T.~T., Smith D.~E., Larson R.~G. and Chu S.,
   Science \textbf{268}, (1995) 83. 
  
\bibitem{PRE_Larson_1997}
  Larson R.~G., Perkins T.~T., Smith D.~E. and Chu S.,
   Phys. Rev. E \textbf{55}, (1997) 1794. 
  

\bibitem{Macromol_Brochard_1995}
  Marciano Y. and Brochard-Wyart F.,
   Macromolecules \textbf{28}, (1995) 985-990. 
  
\bibitem{JPCB_Rowghanian_Grosberg_2011}
  Rowghanian P. and Grosberg A.~Y.,
   J. Phys. Chem. B, (2011). 
  


  
\end{thebibliography}
\end{document}